\input{aipcheck}

\documentclass[
    ,final            
  ]
  {aipproc}

\layoutstyle{6x9}

\begin{document}

\title{A domain-wall-brane-localized standard model}


\author{Raymond R. Volkas}{
  address={School of Physics, Research Centre for High Energy Physics, The University of Melbourne, Victoria 3010, Australia}}

\begin{abstract}

We sketch an SU(5) bulk model in $4+1$-dimensions that plausibly produces an 
effective $3+1$-dimensional standard model dynamically-localized on a domain wall.

\end{abstract}

\maketitle


\section{Introduction}

It has long been speculated that our universe might be a domain wall brane embedded in a background
$4+1$-dimensional spacetime \cite{Rub&Shap} (see also \cite{ADD, RS1, RS2, Antoniadis, Antoniadisetal, Akama,
Visser, Gibbons&Wiltshire}).  
The main challenge in constructing such a model is the simultaneous dynamical
localization of fermions, gauge bosons, gravitons and Higgs bosons.  In this talk, I sketch how various
mechanisms for the localization of these four classes of particle can be assembled into a coherent
model realizing the standard model (SM), or something close to it, as the effective $3+1$-dimensional
theory for the localized fields \cite{DavGeoVolk}.

The main aesthetic motivation is to treat the extra dimension on an equal footing to the usual
spatial dimensions in the action, with the distinction arising only at the
solution level.  Since in such a theory the brane is not fundamental but rather a soliton
solution, it also has the advantage of answering the question: what is the brane made of?  (The
answer is scalar fields.)  Finally, it seems more satisfying to provide dynamical localization
mechanisms rather than positing {\em ab initio} that certain fields are localized to a
fundamental brane.  

The theory we shall describe \cite{DavGeoVolk} requires an SU(5) gauge symmetry in the bulk, spontaneously
broken to SU(3) $\otimes$ SU(2) $\otimes$ U(1) on the wall.  It has two immediate advantages over
$3+1$-dimensional SU(5) models: absence of the tree-level $m_e = m_d$ mass relations, and the
chance to naturally suppress colored-Higgs-induced proton decay.

\section{The model}

Let us start with gauge boson localization.  The best understood proposal is that due to
Dvali and Shifman \cite{DvaliShifman}: Consider a theory with gauge group $G$ containing an adjoint scalar multiplet $\chi$
plus a singlet $\eta$ with $\eta \to -\eta$ imposed as a discrete symmetry.  Arrange the Higgs potential
parameters so that a solution of the coupled Klein-Gordon equations has $\eta$ forming a kink
as a function of the extra-dimensional coordinate $y$, with the relevant component of $\chi$ assuming
a $1/\cosh$-like configuration.  This causes the spontaneous breaking $G \to H$ inside the domain
wall, with $G$ rapidly restored in the bulk.  Taking the bulk theory to be asymptotically-free and
confining, we deduce that the $H$ gauge bosons will be localized to the wall.  The point is that these
gauge bosons propagate as massless gauge particles (or relatively light $H$ glueballs)
inside the wall, but as constituents of massive
$G$ glueballs in the bulk.  The mass gap plausibly ensures their localization.  Alternatively, one may view
the bulk as a dual superconductor, and observe that the field lines of a $G$-charge source in the wall
will be repelled from the (smeared out) boundary with the bulk, leading to effective $3+1$-dimensional behavior
in the large distance limit (within the wall) \cite{A-H&Schmaltz2, RubaDub}.  
For our application, $G$ is SU(5) and $H$ is the SM
subgroup.  It is interesting that gauge boson localization immediately motivates grand unification
(in the bulk).

It can be readily verified that the Higgs potential
\begin{equation}
V_{\eta\chi} = (c\eta^2 - \mu_\chi^2){\rm Tr}(\chi^2) + \lambda_1 \left[{\rm Tr}(\chi^2)\right]^2
 + \lambda_2 {\rm Tr}(\chi^4) + l(\eta^2 - v^2)^2,
\label{eq:Vetachi}
\end{equation}
admits the required solution-type for a large range of parameters.  Generally, the solutions can only
be obtained numerically, but the simple analytic form
$\eta(y) = v\tanh(ky)$, $\chi_1(y) = A\, {\rm sech}(ky)$
follows when certain relations on the Higgs parameters are enforced, with the inverse width $k$ and
the amplitude $A$ given in terms of those parameters.  The notation $\chi_1$ signifies the component
of the SU(5) adjoint that induces breakdown to the SM gauge group.
In writing the above quartic potential, 
we adopt the view that our non-renormalizable $4+1$-d model is but an effective theory, valid below
an ultraviolet cutoff, and we analyse here only the dominant, lowest-dimension operators.  This
Higgs potential, and all subsequent Lagrangians, are truncations.

We now add the fermions $\Psi_5 \sim 5^*$ and $\Psi_{10} \sim 10$, and couple them to $\eta$
and $\chi$:
\begin{equation}
Y_{DW} =  h_{5\eta} \overline{\Psi}_5 \Psi_5 \eta +  h_{5\chi} \overline{\Psi}_5 \chi^T \Psi_5
 +  h_{10\eta} {\rm Tr}(\overline{\Psi}_{10} \Psi_{10})\eta 
- 2 h_{10\chi} {\rm Tr}(\overline{\Psi}_{10} \chi \Psi_{10}).
\label{eq:YDW}
\end{equation}
We solve the Dirac equations,
\begin{equation}
 \left[ i\Gamma^M \partial_M - h_{n\eta} \eta(y) - \sqrt{\frac{3}{5}}\frac{Y}{2}\, h_{n\chi} \chi_1(y) 
\right] \Psi_{nY}(x,y) = 0,
\label{eq:Dirac}
\end{equation}
in the $\eta(y)$, $\chi_1(y)$ background,
where $n=5,10$ and $Y$ is the weak-hypercharge of the SM components denoted $\Psi_{5Y}$ and $\Psi_{10Y}$.
We seek separated-variable solutions, $\Psi_{nY}(x,y) = \psi_{nY,L}(x)f_{nY}(y)$, where the $\psi$'s are 
left-chiral ($\gamma_5 \psi_L = -\psi_L$) $3+1$-d fields obeying the massless $3+1$-d Dirac equation.
The localization profiles are then
\begin{equation}
f_{nY}(y) \propto e^{-\int^y b_{nY}(y') dy'}.
\label{eq:fnY}
\end{equation}
where the zeroes of 
\begin{equation}
b_{nY}(y) \equiv h_{n\eta} \eta(y) + \sqrt{\frac{3}{5}}\frac{Y}{2}\, h_{n\chi} \chi_1(y)
\label{eq:bnY}
\end{equation}
are the localization centers.  The various SM components are {\em split} \cite{A-H&Schmaltz}
along $y$ due to the
different linear combinations of $\eta$ and $\chi_1$ that they feel.  Figure 1 shows examples
of such profiles.

\begin{figure}
  \includegraphics[height=.22\textheight]{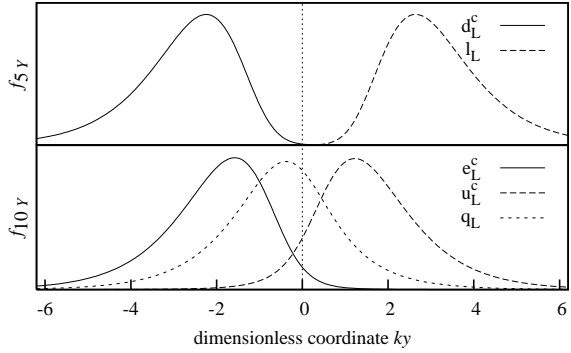}
  \caption{Typical fermion localization profiles normalized so they square-integrate to one.}
\end{figure}

Electroweak symmetry breaking can now be incorporated by introducing the scalar $\Phi \sim 5^*$,
which couples as per
\begin{eqnarray}
Y_5 & = & h_{-} \overline{(\Psi_5)^c} \Psi_{10} \Phi + h_{+} \epsilon^{ijklm} {\overline{(\Psi_{10})^c}}_{ij} {\Psi_{10}}_{kl} \Phi^{*}_m + h.c.\\
V_{\rm rest} & = & \mu_\Phi^2 \Phi^{\dagger} \Phi + \lambda_3 (\Phi^{\dagger} \Phi)^2
+ \lambda_4 \Phi^{\dagger} \Phi \eta^2
+ 2 \lambda_5 \Phi^{\dagger} \Phi {\rm Tr}(\chi^2)\nonumber\\
& +  & \lambda_6 \Phi^{\dagger} (\chi^T)^2 \Phi + \lambda_7 \Phi^{\dagger} \chi^T \Phi \eta.
\label{eq:Vrest}
\end{eqnarray}
The field $\Phi$ contains the electroweak Higgs doublet $\Phi_w$ and a colored scalar $\Phi_c$.  Writing
both in separated variable form, $\Phi(x,y) = p(y)\phi(x)$, where the $\phi$'s are required to satisfy 
$3+1$-d massive Klein-Gordon equations, the $4+1$-d KG equations yield
\begin{equation}
-\frac{d^2}{dy^2} p_{w,c}(y) + W_Y(y) p_{w,c}(y) = m^2_{w,c} p_{w,c}(y),
\label{eq:pwc}
\end{equation}
with a weak-hypercharge-dependent effective potential,
\begin{equation}
W_Y(y) = \mu_\Phi^2 + \lambda_4 \eta^2 + \lambda_5 \chi_1^2
    + \frac{3 Y^2}{20} \lambda_6 \chi_1^2 + \sqrt{\frac{3}{5}} \frac{Y}{2} \lambda_7 \eta \chi_1.
\label{eq:pwcpot}
\end{equation}
with $m^2_{w,c}$ being the $3+1$-d squared masses.  The two effective potentials can be arranged to
produce $m_c^2 > 0$, $m^2_w < 0$ and localized $\phi$'s.  The tachyonic $m^2_w$ triggers
spontaneous electroweak symmetry breaking on the brane.  Figure 2 displays suitable potentials.

\begin{figure}
  \includegraphics[height=.22\textheight]{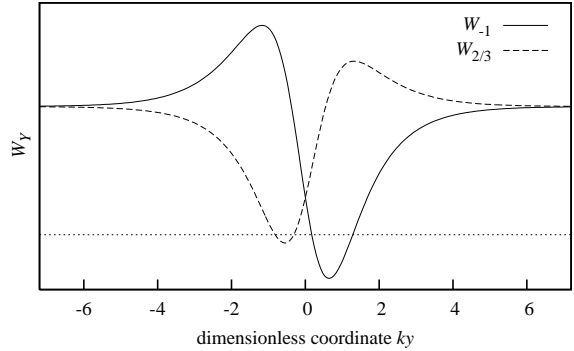}
  \caption{Sample Higgs and colored scalar localization potentials.  The horizontal
line is $W=0$, and $W_{-1}$ has a negative eigenvalue $m^2_w$ triggering electroweak
symmetry breakdown.}
\end{figure}

The usual SU(5) mass relations $m_e = m_d$ are {\em not} obtained from $Y_5$ because the
$3+1$-d masses depend on overlap integrals of bulk profiles which differ among the SM
components.  Similarly, one may be able to suppress colored-scalar-induced proton decay 
because the relevant overlap integrals are small due to the splittings.

The above analysis had gravity switched off.  We now include Einstein-Hilbert and bulk
cosmological terms.  By fine-tuning the latter, one may localize the graviton by the
smooth domain-wall-brane version of type-2 Randall-Sundrum \cite{RS2} (the warp factor exponent
behaves something like $-\log(\cosh ky)$ rather than $-|ky|$).  The various effective
Schr\"{o}dinger potentials that induce localization now change in character: they
get warped-down to zero asymptotically in the bulk, rather than approaching nonzero constants.  
This means that mode-continua begin at zero mass for all particles, an effect
well-known for gravitons in type-2 RS.  What were previously excited bound (i.e.\ localized) states
now become {\em resonances}.  The non-resonant modes are greatly suppressed in amplitude
on the brane, because the wave-functions have to tunnel through potential barriers.
Thus, despite the continua starting at zero mass, an effective $3+1$-d theory is
still produced (see Ref.~\cite{DavGeo} for more details), retaining the general character
of the flat-space toy model.

We conclude by saying that the above construction plausibly produces an effective wall-localized
theory that is close to being the standard model.


\begin{theacknowledgments}
This work was performed in collaboration with Damien P. George and Rhys Davies \cite{DavGeoVolk}.  
It was supported by the Australian Research Council.
\end{theacknowledgments}



\bibliographystyle{aipproc}   

\bibliography{references}

\IfFileExists{\jobname.bbl}{}
 {\typeout{}
  \typeout{******************************************}
  \typeout{** Please run "bibtex \jobname" to optain}
  \typeout{** the bibliography and then re-run LaTeX}
  \typeout{** twice to fix the references!}
  \typeout{******************************************}
  \typeout{}
 }

\end{document}